# Wild, Wild Wikis: A way forward

Robert Charles[1] and Adigun Ranmi[2]

[1]*Laboratoire Lorrain de recherche en informatique et ses applications*
*Campus Scientifique - BP 239  54506 Vandoeuvre-lès-Nancy Cedex, France*
*Abiodun-charles.robert@loria.fr*

[2]*Department of Computer Technology Yaba college of Technology yaba – Nigeria*
*ranmiadigun@yahoo.com*

## Abstract

*Wikis can be considered as public domain knowledge sharing system. They provide opportunity for those who may not have the privilege to publish their thoughts through the traditional methods. They are one of the fastest growing systems of online encyclopaedia. In this study, we consider the importance of wikis as a way of creating, sharing and improving public knowledge. We identify some of the problems associated with wikis to include, (a) identification of the identities of information and its creator (b) accuracy of information (c) justification of the credibility of authors (d) vandalism of quality of information (e) weak control over the contents. A solution to some of these problems is sought through the use of an annotation model. The model assumes that contributions in wikis can be seen as annotation to the initial document. It proposed a systematic control of contributors and contributions to the initiative and the keeping of records of what existed and what was done to initial documents. We believe that with this model, analysis can be done on the progress of wiki initiatives. We assumed that using this model, wikis can be better used for creation and sharing of knowledge for public use.*

## 1. Introduction

Contributions to public domain initiatives can be very important in aggregation of public knowledge. Some of the public domain resources include ebooks, free music / videos, public forums, free images and public domain encyclopaedias. We can easily identify two types of public domain resources: static resources which do not change with time. Ebooks, music, images can be seen as static whereas public forums and encyclopaedias that changes with time particularly when they are shared using the internet resources are dynamic resources. In this paper, we are interested in dynamic public domain resources. We can identify several examples in this category. wikis, flickr and del.icio.us are well examples of dynamic public resources. Our particular attention is on wikis. The first question that comes to mind is what is wiki? Stafford and Webb defined wiki "*as website where users can add, remove, and edit every page using a web browser*".

In this study, we see wiki as one of the methods of aggregating public knowledge on the internet. Multiplicity of wikis is not a problem but an indication of its usefulness and popularity. The objective of this paper is to see how contributions in public wikis can be enhanced using an annotation model. Any contributed knowledge will normally attract criticism particularly when it is in public domain We can refer to three scenarios when an individual consults a public domain wikis. (a) He may feels that something is wrong with the content of the information (b) He may feel that the information contain therein is accurate and justifiable (c) He may feel that the information is good but can be improved upon. In the last two situations, the user may decide to improve the quality of information in the system by adding his information to negate or improve what was initially available. The question is, how can he improve the quality of information in a wiki?

## 2. Public resources: wonders . worries .wrangles

Wikis are not personal or public blogs. A wiki is a type of website that allows users to easily add, remove, or otherwise edit and change most content of the web resources, sometimes without the need for registration. This ease of interaction and operation makes a wiki an effective tool for collaborative writing. The term wiki can also refer to the collaborative software itself (wiki engine) that facilitates the operation of such a website or to certain specific wiki sites, including the computer science site (and original wiki), WikiWikiWeb, and the online encyclopedias such as Wikipedia. The use of wikis divers. Eric Baldeschwieler, director of software development of Yahoo inc, was quoted among others by Twik[1] and David Warlick's CoLearners[2] on 12 August 2004, to have said that "Wiki is being used by Yahoo to internally manage documentation and project planning for their products". Gloria McConnell [7], enumerated the following uses of wikis:
- internal message board,
- software or documentation archive,
- broader collaboration,
- tracking issues ("bugs") and features,
- software design and documentation,
- knowledge base or FAQ system,
- company intranet.

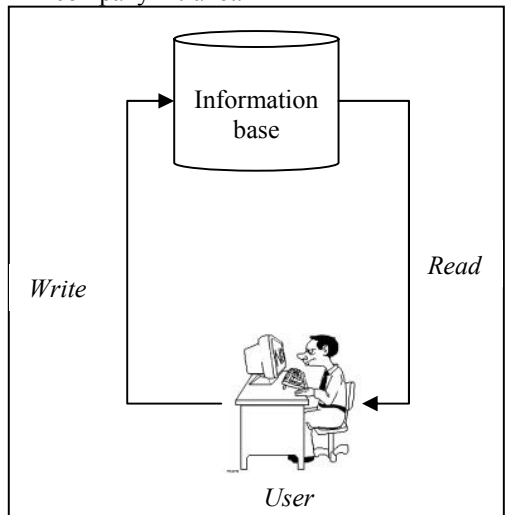

**Figure 1:** Process of participation in wikis

---

[1] http://twiki.org/cgibin/view/Main/TWikiSuccessStoryOfYahoo
[2] http://davidwarlick.com/wiki/pmwiki.php?n=Main.DougMelillo

To demonstrate the importance and the acceptance of wikis as a method of collaborative writing, some major programming languages are introducing patches, plugins, modules or API to make creation of public domain resources (like wikis) easy. TCL/TK introduced what they call wikit which can be implemented through CGI or standalone. Javapedia project is an implementation of wikis in java with such classes as WebChanges, WebIndex, WebStatistics, WebPreferences, CreateANewPage etc…

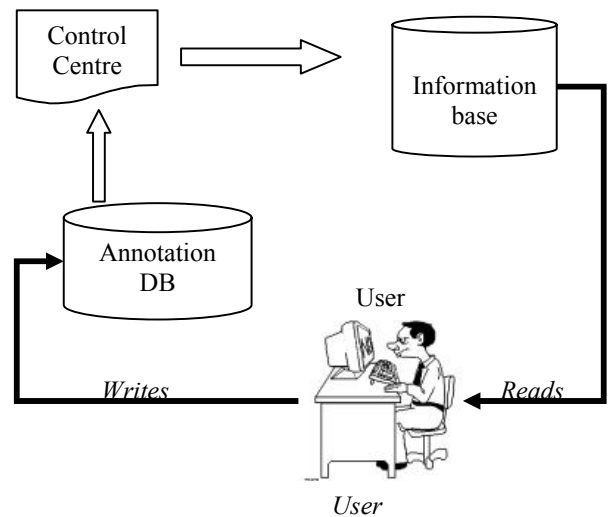

**Figure 2:** Process for improvement in wikis

We can identify three distinctive types of wiki web sites that encourage collaborative writings. The first type can be qualified as public and open wikis. In this case, the participation is open to the public and the subject of discussion is open-ended e.g. wikipedia.org. The second type are the local wikis that are built around local users on local intranet or web servers that requires personal account on the wiki systems server, example of these includes http://wiki.loria.fr. The third type can be classified as "specific product wikis". In this case, the wikis were initiated to encourage discussions around a product instead of the use of blogs. These discussions can be seen as a kind of feedback on these products. Some of these include, wiki.mozilla.org, VBWiki, FoxWiki, SQLWiki, etc. Some of the inherent benefits of wikis and other public domain resources include:

- Ease of public access to information: because domain resources like wikis are directly accessible (for reading and editing) without restraints in form of security and pre-requisite, it is one of the fastest means of accessing public knowledge.
- "Freedom of speech": because of the free access to publish personal opinions on issues, individuals can publish their ideologies on issues without seeking for "authorized publishers".
- Growth: as a result of the freedom and the ease of information access, wikis are one of the fastest growing online encyclopaedias. The growth makes it a formidable source of information.

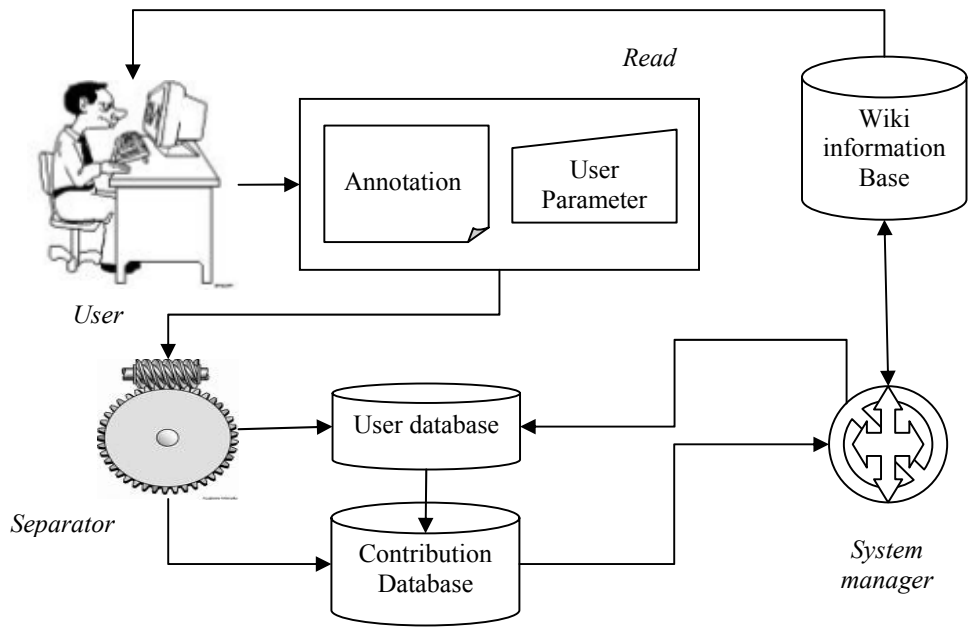

**Figure 3**: An architecture for wiki improvement

With all these inherent benefits, several problems were identified with this means of knowledge sharing [4]. Some of the basic problems associated with wikis can be summarized as:

- The information therein lacks authenticity because they may not necessarily have trade marks in terms of author and other identities to identify information.
- Sometimes, useful or more authentic information are overwritten by less important or less accurate information.
- There are no known ways to justify the credibility of authors in wikis.
- Vadalism: There is the possibility to deliberately delete, change content of information to reduce the quality of information in wiki.
- Authority: The authority over the contents in wikis is weak. This makes it a sort of public garbage.

## 3. A view for wholesome.worthy.wikis

Private wikis or local wikis may not suffer some of the problems enumerated in the previous section because their coverage is local and there acceptability is narrower. For a wiki to be popular, and be more useful, it must cut across wider domain and wider audience. When this happens, the authenticity and the quality becomes a problem. In this case, we advocate some kind of moderation to assure its usefulness. The moderation process can be compared to what the industries called "Quality control". In order to achieve this moderation, we approach the problem based on some propositions from an annotation model called AMIE (Annotation model for information exchange). AMIE was originally conceived for annotation in information research for decision making [10]. The basis of AMIE is the fact properties of document, with properties of an annotation creator in time are sufficient to qualify set of annotation for decision making. In this case, we base our definition on the same premises with that in [10]. It states that "Documents are traces of human activities" [8]. We assume that wikis are traces of human activities. If this is the case, wikis can be considered as documents just as other forms of traditional documents. We base our propositions for improved wikis on the following assumptions:

- Readers or writers of wikis are users of wikis.
- There are no anonymous writers of wikis
- Wikis are documents with specific parameters
- Newly created wikis are annotations
- Wikis are created with reference to time and their creators
- Annotations have specific characteristics

We will define annotation as *"additional explanation to a document section to serve as definitions, examples, references, etc"*. In essence, annotation is additional information to existing one.

In this case, additions to wikis can be seen as annotation. In the light of previous works done in annotation [1][2][3][10], additions to wikis are annotation to existing document. We also know that existing wikis are the source document necessary in all annotations processes [5]. In our proposition, we encourage that all "proposed new contributions" should be sent to an annotation database for control before it can be added to the information database as indicated in the Figure 2. Here we use the term "proposed new contributions" to include all possible actions in wikis. Before the sending of "new propositions", the quality of what is send must be determined. In our advancement for quality wikis, some pre-requisites are expected during its creation. These requirements can include (a) a well labelled contribution including specific information relating to the contribution, for example if the contribution is for replacement, edition, or suggestion to remove existing information if the existing information is obviously wrong, obsolete or questionable and (b) the identity of the user/creator.

## 4. A synopsis for improvement

The architecture of a system that can improve the quality of information in wiki systems may consist of three groups of processes and three databases. The first group of processes emanates from the user who makes a formal request from a wiki information base. He is interested in a particular kind of information. We are not interested in the method and syntax of his request here. We expected that, a user of any system based on our model should not be anonymous. He is expected to send his users parameters along with his request for information. One of the popular ways of confirming user's information from an information system during information research is through login. Whatever algorithm that may be used in responding to his request is not our concern in this study. We are satisfied with the fact that he can receive specific information from the system. In the normal wiki system, (Figure 2), information received from the system is re-processed based on the perceptions of the user (addition, correction, edition, deletion or overwrite). In our own proposition, the output of processes from a user is sent to a separator. It is thought that the input from the user is a combination of user's contribution and his identity. The separator separates the user's parameters from the user's contribution. User's contribution is sent to contribution database with all specificities such as (a) what was the original document referenced and the specific suggestions (contribution) from the user (b) date of contribution (c) identity of creator/user.

We can have an automatic system manager or human manager that checks for new submission in the database of contribution. Our objective in this work is not to provide an algorithm for an automatic checking in such system.

What is submitted can be checked based on defined criteria. We had defined document as a trace of human activities, it then implies that these trace must be linked to individuals, time, events and space. Realize that anonymous contributions may not be acceptable. It then means that a complete identity of an author is a requisite in the database of author that is linked to the shared knowledge. The criteria to check may include the author's name and identity. It may also be linked to other databases so evaluate the claims of the author. For example it the author has previous publications related to his contribution. Realize that the author is identified by his affiliation and experiences in the area of his contribution. He may be a regular contributor to the forum therefore his identity is verified from the user database. The system manager may also verify if there are other publications related to the contribution in other databases; in that case other checks may be necessary. The algorithm for checking may be much longer, extensive and comprehensive.

It is after these "checking" that contribution may be added to the information data base. The system manager should also provide ways of reversing a contribution. It should be noted that this control enable the manager to keep record of the evolutions in wikis for possible analysis.

## 5. Conclusion and perspective

We have demonstrated the fact that wiki systems, (particularly public wikis) lack quality and authenticity. These problems may be as a result of the fact that information is expected to be accessible as fast as possible. We also enumerated other problems associated with public domain resources and particularly wikis. In other to solve some of these problems, we proposed a model that can be used to add information to wiki systems without necessarily compromising the quality of this information. It is expected that in no distant time, we will be able to implement these propositions.